\documentclass[]{elsart}

\pdfoutput=1

\usepackage{epsfig}
\usepackage{graphics}
\usepackage{graphicx}
\usepackage[centertags]{amsmath}
\usepackage{amsfonts}
\usepackage{amssymb}
\usepackage{url}
\usepackage{subfigure}

\begin{document}

\begin{frontmatter}

\title{Programmable reconfiguration of \\ \emph{Physarum} machines}

\author{Andrew Adamatzky \and Jeff Jones} 

\address{University of the West of England, Bristol BS16 1QY, United Kingdom \\ \{andrew.adamatzky,jeff.jones\}@uwe.ac.uk}

\date{}

\maketitle

\begin{abstract}
\noindent
Plasmodium of \emph{Physarum polycephalum} is a large cell capable of solving  graph-theoretic, optimization 
and computational geometry problems due to its unique foraging behavior. Also the plasmodium is unique biological 
substrate that mimics universal storage modification machines, namely the Kolmogorov-Uspensky machine. In the plasmodium 
implementation of the storage modification machine data are represented by sources of nutrients and memory structure by protoplasmic tubes connecting the sources. In laboratory experiments and simulation we demonstrate how the plasmodium-based 
storage modification machine can be programmed. We show execution of the following operations with active zone (where computation
occurs): merge two active zones, multiple active zone, translate active zone from one data site to another, direct active zone.
Results of the paper bear two-fold value: they provide a basis for programming unconventional devices based on biological substrates
and also shed light on behavioral patterns of the plasmodium.

\vspace{0.5cm}

\end{abstract}

\end{frontmatter}

\section{Introduction}

\emph{Physarum polycephalum}\footnote{Species of order \emph{Physarales}, subclass \emph{Myxogastromycetidae}, 
class \emph{Myxomecetes}, division \emph{Myxostelida}}, commonly known as a true or multi-headed slime mold, 
is --- at one stage of its complicated life cycle -- a single cell with many diploid nuclei. 
This syncytial mass of protoplasm, called plasmodium, looks like 
amorphous yellowish mass.  The plasmodium behaves and moves as a giant amoeba. It feeds on bacteria, spores and other microbial creatures. When foraging for its food the plasmodium propagates towards sources of food particles, surrounds them, secretes enzymes and digests the food.  Typically the plasmodium forms a congregation of protoplasm in a food source it occupies. 
When several sources of nutrients are scattered in the plasmodium's range, the plasmodium forms a network of protoplasmic tubes
connecting the masses of protoplasm at the food sources. When we think of food sources as nodes and protoplasmic tubes as edges, we say the plasmodium develops a planar graph.   

Nakagaki \emph{et al}~\cite{nakagaki_2000,nakagaki_2001,nakagaki_2001a} showed that the topology of the  plasmodium's protoplasmic 
network optimizes the plasmodium's harvesting on distributed sources of nutrients and makes more efficient flow
and transport of intra-cellular components. Therefore the plasmodium is considered as a parallel computing substrate 
complementary~\cite{adamatzky_naturewissenschaften_2007} to existing massive-parallel reaction-diffusion 
chemical processors~\cite{adamatzky_2005}. Experimental observations suggest that during development of its protoplasmic network the plasmodium undergoes transitions between various classes of proximity graphs~\cite{adamatzky_ppl_2008}. It starts with spanning 
trees~\cite{adamatzky_kybernetes_2007} and may complete its protoplasmic network 
development as a Delaunay triangulation~\cite{shirakawadelone}.

Implementation of a general purpose computing machine is the most remarkable feature of the plasmodium of \emph{ Physarum 
polycephalum}. In~\cite{adamatzky_ppl_2007} we experimentally demonstrated that the plasmodium
can implement the Kolmogorov-Uspensky (KUM) machine~\cite{kolmogorov_1953,uspensky_1992}, a mathematical 
machine in which the storage structure is an irregular graph. The KUM is a forerunner and direct
`ancestor' of Knuth's linking automata~\cite{knuth_1968}, Tarjan's reference machine~\cite{tarjan_1977}, and 
Sch\"{o}nhage's storage modification machines~\cite{schonhage_1973,schonhage_1980}. The storage modification machines are 
basic architectures for random access machines, which are the basic architecture of modern-day computers. The plasmodium-based implementation of KUM~\cite{adamatzky_ppl_2007} provides a first-ever biological prototype of a general purpose computer.

The key component of the KUM is an active zone~\cite{kolmogorov_1953,uspensky_1992}, which may be seen as a computational
equivalent to the head in a Turing machine. Physical control of the active zone is of utmost importance because it determines
functionality of the biological storage modification machine.

In the present paper we show --- in laboratory and computer experiments with \emph{ Physarum polycephalum} --- how basic operations
{\sc Add node}, {\sc Add edge}, {\sc Remove Edge} are implemented in the Physarum machine. We also provide unique results on
controlling movement of an active zone.  The paper is structured as follows. We provide a very brief introduction to Kolmogorov-Uspensky machine in Sect.~\ref{KUM}. 
Section~\ref{methods} describes a simple experimental setup for study of plasmodium.  A particle-based model of the plasmodium is
presented in Sect.~\ref{model}. In Sect.~\ref{results} we discuss results of laboratory experiments and computer simulation on reconfiguration of  basic protoplasmic graphs. Overview of the results and future experiments are outlined in Sect.~\ref{discussion}.

\section{Physarum machine}
\label{KUM}

Kolmogorov-Uspensky Machine (KUM) is defined on a labeled undirected graph (storage structure) with bounded degrees of nodes and bounded number of labels~\cite{kolmogorov_1953,uspensky_1992}. KUM executes the following operations on its storage structure:
select an active node in the storage graph; specify local active zone, i.e. the node's neighborhood; 
modify the active zone by adding a new node with the pair of edges, connecting the new node with the active node; delete a node with a pair of incident edges; add/delete the edge between the nodes. A program for KUM specifies how to replace the neighborhood of an active node (i.e. occupied by an active zone) with a new neighborhood, depending on the labels of edges connected to the active node and the labels of the nodes in proximity of the active node~\cite{blass_gurevich_2003}.

\begin{figure}
\centering
\includegraphics[width=0.5\textwidth]{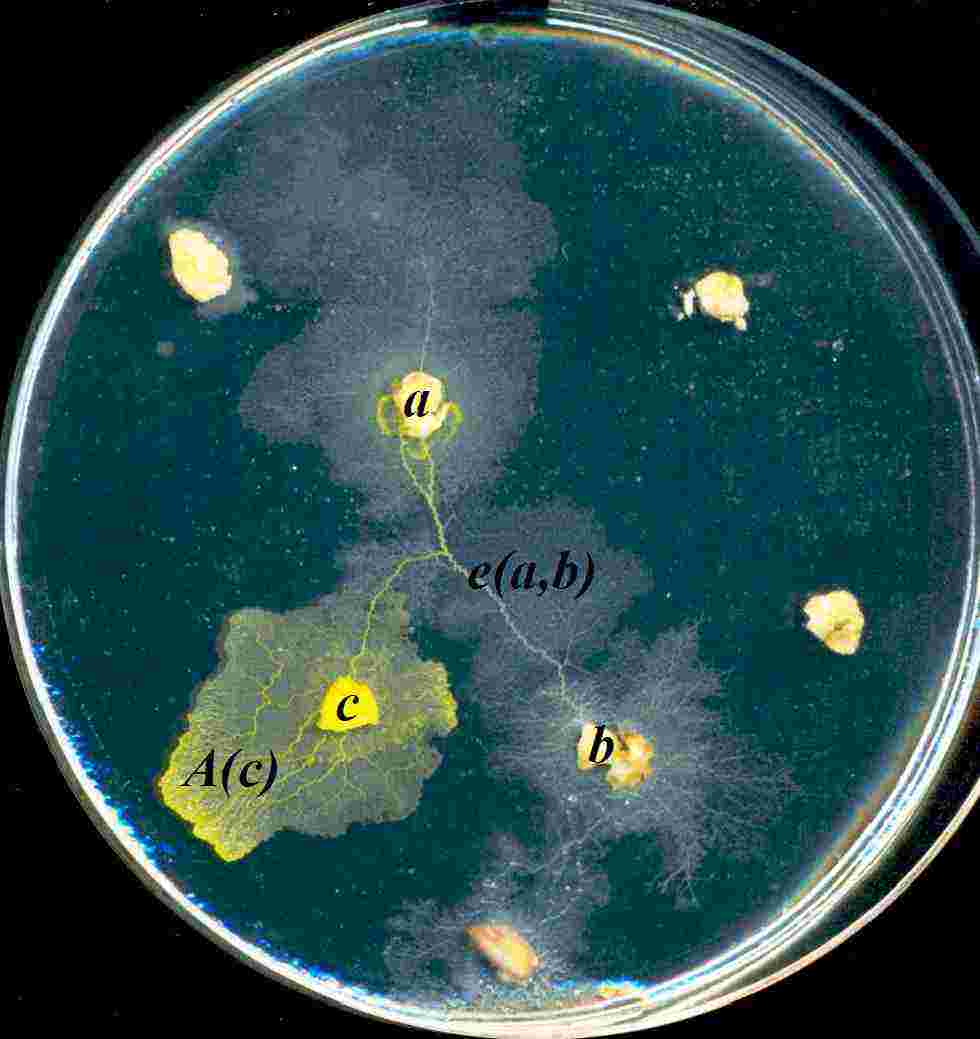}
\caption{An exemplar snapshot of Physarum machine. Protoplasmic tube connecting flakes $a$ and $b$ represents
an edge $e(a,b)$ of KUM. Active zone $A(c)$ emerges in the node $c$.}
\label{examplePM}
\end{figure}

Physarum machine is a biological implementation of KUM, where a node of the storage structure is represented by a source of nutrients (e.g. an oat flake); an edge connecting two nodes is a protoplasmic tube linking two sources of nutrients corresponding 
to the nodes; and, an active zone is domain of space (which may include food sources) occupied by a propagating pseudopodium. An 
sample architecture of Physarum machine is illustrated in Fig.~\ref{examplePM}. 

In Physarum machine the computation is implemented by active zone, or several active zones. To make the computation process
programmable one needs to find ways of sensible and purposeful manipulation with the active zones. Several operations, manipulation
procedures, are discussed in the present paper.

\section{Experimental method}
\label{methods}

The plasmodia of \emph{Physarum polycephalum} were cultured on wet paper towels, fed with oat flakes, and moistened regularly. We subcultured the plasmodium every 5-7 days. Experiments on growing spanning trees were undertaken in standard Petri dishes, 9~cm in diameter. A substrate was a wet filter paper. We preferred the filter paper not 2\% agar gel, because the paper offers less favoured conditions for the plasmodium growth, and thus less branching of the propagating pseudopodia is observed (see details in~\cite{adamatzky_delacycostello_shirakawa_2008,takamatsu_2007}). The Petri dishes
with plasmodia were kept in darkness and only exposed to light during observation and recording of images. 

Data points, to be connected by protoplasmic graphs, were represented by oat flakes. Photographs of the protoplasmic networks, developed by the plasmodium, were made using the digital camera FinePix S6500. Colours of original images were enhanced by increasing contrast to 100 and decreasing gamma correction to 0.3. 

\section{Particle-based model approximating the behaviour of \emph{Physarum} plasmodium}
\label{model}

To provide the computer model approximation of \emph{Physarum} plasmodium active zone computation, the laboratory experiment configurations were used to map synthetic environments for the model organism. The model is based on the simple particle behaviours that were used in~\cite{jones_2008} to generate emergent transport networks. The model is computationally very simple, using distributed local sensory behaviours, but approximates some of the complex phenomena observed in \emph{Physarum}: foraging for food stimuli, spatially distributed oscillations, oscillation phase shifting, shuttle streaming, amoebic movement, network formation, surface area minimisation, and network minimisation behaviours. A full description of the model is in preparation, but an overview specifically in relation to plasmodium behaviour is given below.

The model takes a multi-agent (particle) approach to generate emergent plasmodium behaviours. Movement and internal oscillations of the plasmodium is based upon the collective behaviour of the agent population. The movement of agents corresponds to the flux of sol within the plasmodium. Cohesion of the plasmodium is an emergent surface tension property that arises due to the agent-agent interactions. Directional orientation and movement of the plasmodium active front is generated by coupling the emergent mass behaviours with chemoattraction to local food source stimuli. 

\begin{figure}
\centering
\includegraphics[width=0.99\textwidth]{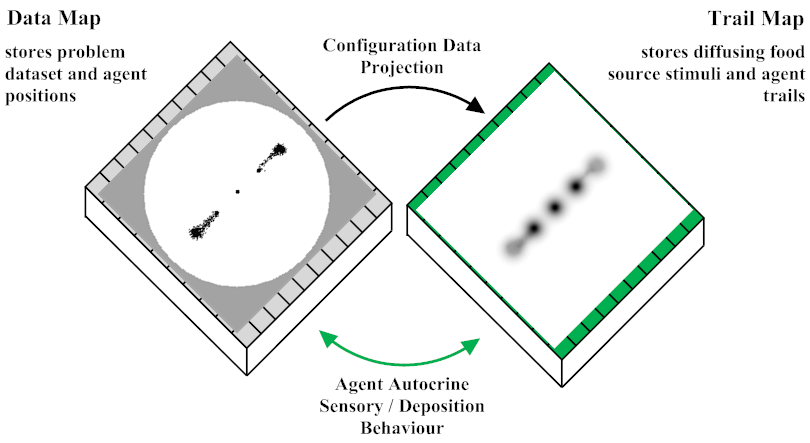}
\caption{Problem data projected to coupled map storing diffusing food source and agent trails.}
\label{pic_coupledmap}
\end{figure}

A population of mobile agents, adopting simple stimulus-response behaviours is created. The population adopts autocrine chemotaxis behaviours --- agents both secrete and sense approximations of chemical trails. The population is coupled to a two-dimensional discrete map representing the problem configuration --- the 'data' map, Fig.~\ref{pic_coupledmap}.

Food source stimuli whose positions are stored in the data map are projected to another coupled map (the 'trail' map) at every step of the scheduler. The strength of the projected food sources can be adjusted with a weighting parameter. When the plasmodium engulfs a food source the stimulus for diffusion is reduced by the encapsulation. This is approximated in the model by damping the food stimulus projection to 1\% of the food value if the site is occupied by an agent. The food stimuli are diffused by means of a simple 3$\times$3 mean filter kernel. The combined diffusion stimuli and agent trails stored in the trail map are degraded at every scheduler step by 0.01 to maintain a relatively steep diffusion gradient away from the food source. The diffusion gradient corresponds to the quality of the nutrient and substrate of the plasmodium's environment (for example the different growth patterns seen in soaked filter paper and agar substrates), and differences in the stimulus strength, stimulus area, affect both the steepness, and propagation distance of the diffusion gradient and affect the growth patterns of the synthetic plasmodium.

\begin{figure}
\centering
\includegraphics[width=0.5\textwidth]{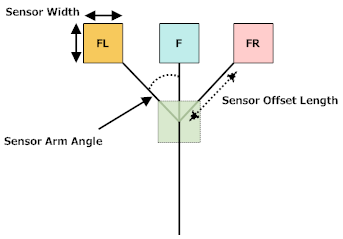}
\caption{Morphology of agent particle.}
\label{pic_alg}
\end{figure}

The agent population (size 2000 agents for all experiments by restructuring of the protoplasmic graph, in last case
we employed 3200 agents) is initialised with random agent positions (on lattice $200\times 200$ sites) and random agent orientations. To reproduce the 'wet' \emph{Physarum} experiments agents were inoculated at particular locations (for example food sources) and new agents are initialised at the front of plasmodium growth according to local occupancy measures. These measures are based on a local window surrounding each agent ($\theta_w$), a minimum occupancy threshold ($\theta_{\min}$), and a maximum threshold ($\theta_{\max}$).  In these results $\theta_w$ = 3$\times$3 window, $\theta_{\min}>0$ and $\theta_{\max} < 4$. The morphology of each agent is shown in Fig ~\ref{pic_alg}. Let $l$, $s$, $r$ be 
concentrations of chemo-attractants as measured by left (FL in Fig.~\ref{pic_alg}), forward (F in Fig.~\ref{pic_alg}, and 
right (FR in Fig.~\ref{pic_alg}) sensors. Each agent 
can be characterised at time step $t$ by its current position $x^t$ and the angle $\alpha^t$ the agent rotates at this time step. At each step of the simulation the rotation angle $\alpha$ is calculated as follows:

$$
\alpha^{t+1}=
\begin{cases}
0, & \text{ if } s^t>l^t \text{ and } s^t>r^t \\
-\psi, & \text{ if }  l^t>r^t \text{ and } l^t>s^t \\
\psi, & \text{ if } r^t>l^t \text{ and } r^t>s^t\\
\text{random}(-\alpha, \alpha), & \text{ otherwise }  \\
\end{cases}
$$

Agent's position is updated as follows: 

$$
x^{t+1}=
\begin{cases}
{\bf R}(\alpha)\circ {\bf T}(\delta_x), & \text{ if } \gamma({\bf R}(\alpha)\circ {\bf T}(\delta_x))=0 \\
{\bf R}(\text{random(360)}), & \text{ otherwise } 
\end{cases}
$$

where ${\bf R}(\alpha)$ is rotation on angle $\alpha$ operation and ${\bf T}(\delta_x)$ is a translation
on distance $\delta_x$ operation, and for any Euclidean point $y$ $\gamma(y)=1$ if there is 
an agent residing at site $y$; $\gamma(y)=0$ otherwise. 

The agents sense the concentration of stimuli in the trail map from three forward oriented sensors. At each scheduler step the agents orient themselves towards the strongest trail source by rotating left or right, depending on the source of the stimuli.

For these simulations sensor width is one pixel, sensor arm angle $\beta= 30$ degrees (Fig.~\ref{pic_alg}), $\alpha=45$ degrees, 
sensor offset distance is nine pixels, and the distance moved per step, $\delta_x$ is one pixel. There is significant interplay between the sensor arm angle ($\beta$) and the rotation angle ($\alpha$) parameters. When the two parameters are equal (for example, both 45 degrees), the effect on network formation is to contract the initial network that emerges. If $\alpha<\beta$ the contraction behaviour is increased. If, however, $\alpha>\beta$ then spontaneous branching will appear during the network formation. The branching occurs because the larger rotation angle places an agent's sensory apparatus outside the zone of the trail that the agent was following. The $\alpha$/$\beta$ parameters can be used to tune the network behaviour towards minimisation or towards high connectivity. After the sensory stage every agent attempts to move forward in its current direction (represented by an internal state from 0-360 degrees). If the new site is not occupied, the agent moves to the new site and deposits trail onto the trail map at the new location (deposition value 5 units). If the agent cannot move forwards successfully then no deposition is made to the trail map. 

Inertial movement for each agent is provided by maintaining a floating point representation of the current position, as well as the discrete position corresponding to the image structure.  This effectively allows the agents to 'slide past' one another even when the next cell is occupied (the occupation can only actually happen when a cell becomes free). The inertial behaviour results in the emergence of surging movements in the population and corresponds to the spatial oscillations seen in \emph{Physarum}. The strength of the oscillations can be reduced by firing a 'change direction' event with probability 0.05 (in these results). When the event is triggered for an agent, the floating point position is restored to the discrete position and a new direction is randomly selected. The effect is to dampen the surge of movement caused by the inertial behaviour.

The list comprising the agent population is traversed in a random order for both the sensory and movement stages to avoid any influence from sequential positional updates. In the current version of the model the maximum population size is manually assigned, based on a proportion of the size of the problem configuration area. Work is in progress to automatically assign the population size in relation to food source availability. 

The experiments on the model plasmodium were designed and configured to reflect the `wet' experiments reported in this report. The results images show the experimental field within the confines of the `dish'. The food sources correspond to the oat flake positions used in the `wet' experiments and are indicated as small dark spots. The position of the plasmodium is indicated as the rough textured masses. Evolution of the model for each experiments proceeds from left to right.

\section{Results: Manipulating active zones}
\label{results}

\begin{figure}
\centering
\subfigure[$t$=0~h]{\includegraphics[width=0.4\textwidth]{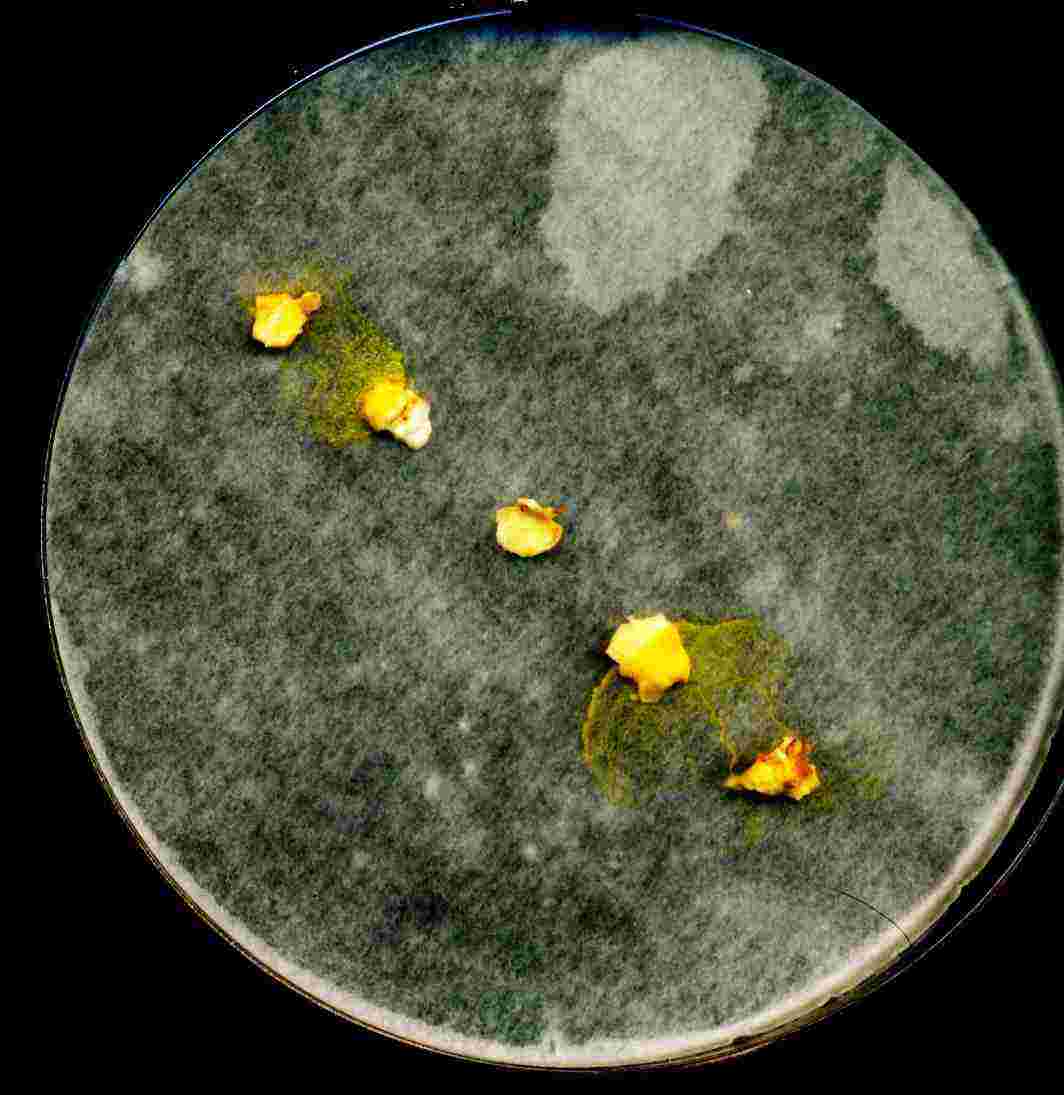}}
\subfigure[$t$=12~h]{\includegraphics[width=0.4\textwidth]{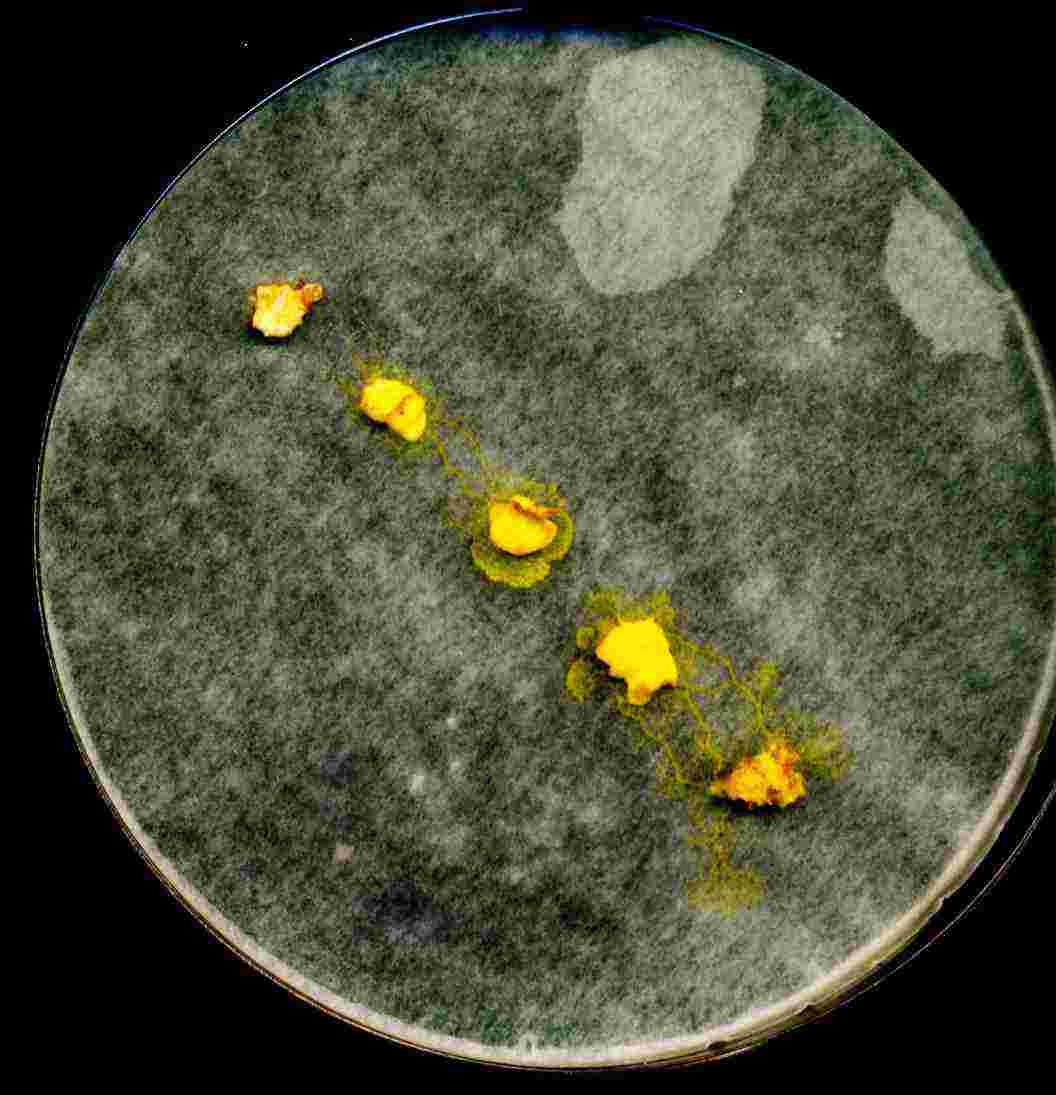}}
\subfigure[$t$=24~h]{\includegraphics[width=0.4\textwidth]{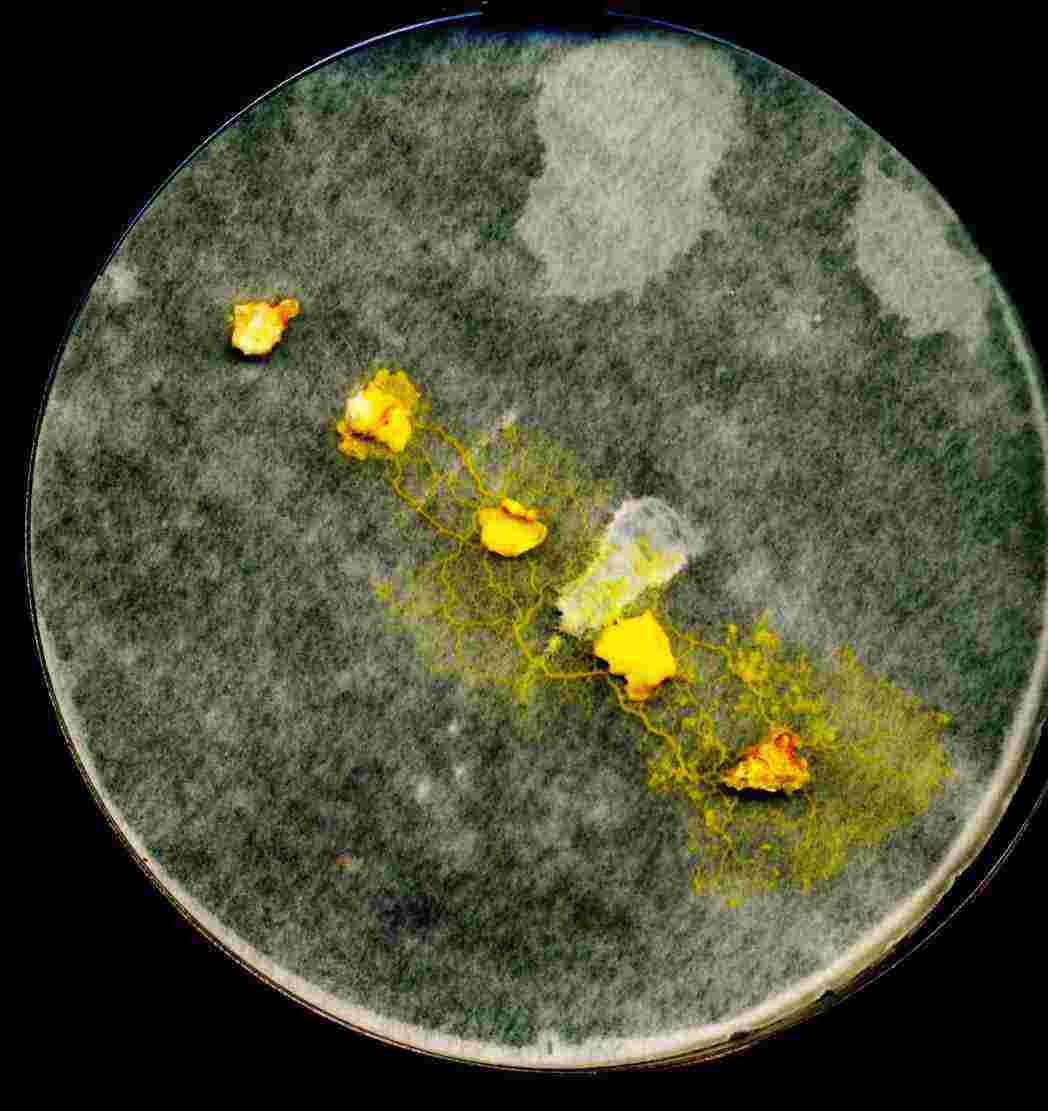}}
\caption{Fusing of active zones. }
\label{fusing}
\end{figure}

The operation {\sc Fuse}$(A_1, A_2)$, fusing of two active zones 
$A_1$ and $A_2$ can be implemented via collision of the active zones. 
An example is demonstrated in Fig.~\ref{fusing}.

Five oat flakes were arranged in a line on a humid filter paper. Two pieces of plasmodium were placed
nearby the extreme flakes, one piece near South-West flake, another piece near North-East. The active 
zones are formed as pseudopodia and propagate towards the center of the chain (Fig.~\ref{fusing}ab). When 
the active zones $A_1$ and $A_2$ collide they fuse and annihilate, {\sc Fuse}$(A_1,A_2)=\emptyset$~(Fig.~\ref{fusing}c). 
Depending on the particular circumstances the new active zone (the result of fusing) may become 
inactive, transform to protoplasmic tubes, or remain active, {\sc Fuse}$(A_1,A_2)=A$, and continue propagation in a new direction.

\begin{figure}
\centering
\subfigure[$t=63$~steps]{\includegraphics[width=0.4\textwidth]{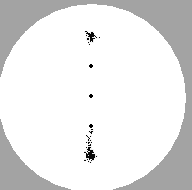}}
\subfigure[$t=105$~steps]{\includegraphics[width=0.4\textwidth]{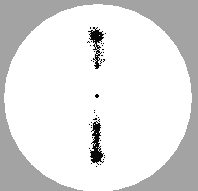}}
\subfigure[$t=147$~steps]{\includegraphics[width=0.4\textwidth]{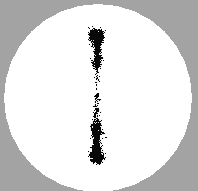}}
\subfigure[$t=294$~steps]{\includegraphics[width=0.4\textwidth]{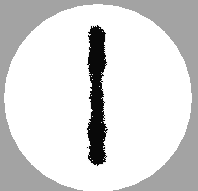}}
\caption{Evolution of model, simulating fusing of active zones. }
\label{fusingModel}
\end{figure}

The model output (Fig.~\ref{fusingModel}) shows the inoculation and multiplication of the plasmodium at the outer nodes. The growing plasmodium is attracted to the inner food sources and the propagation continues inwards from each direction. The plasmodium fuses and maintains its surface area spanning the array of nodes. The model plasmodium was seen to periodically oscillate it's position over the array whilst maintaining coverage.

\begin{figure}
\centering
\subfigure[$t=$0]{\includegraphics[width=0.4\textwidth]{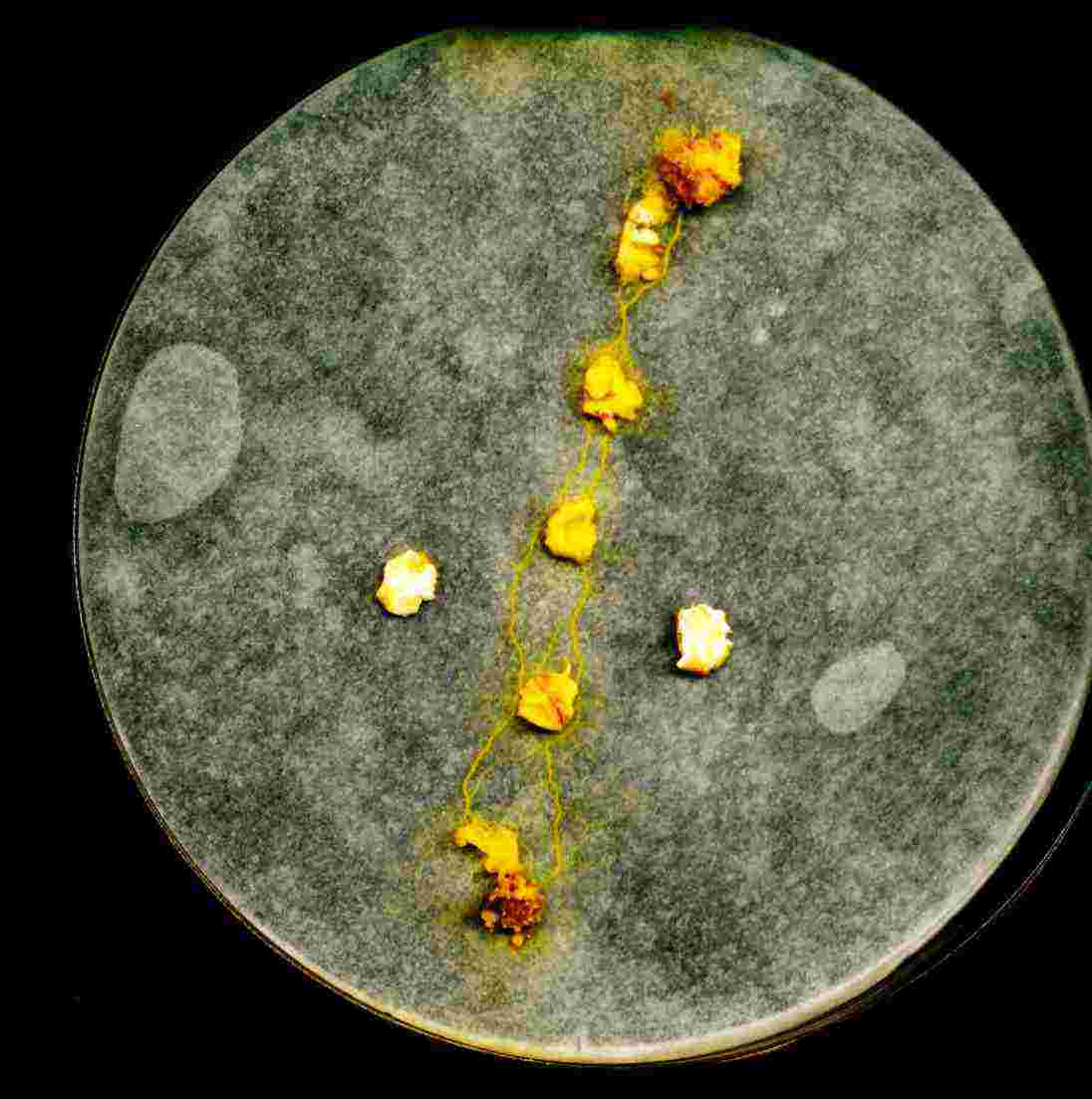}}
\subfigure[$t=$10~h]{\includegraphics[width=0.4\textwidth]{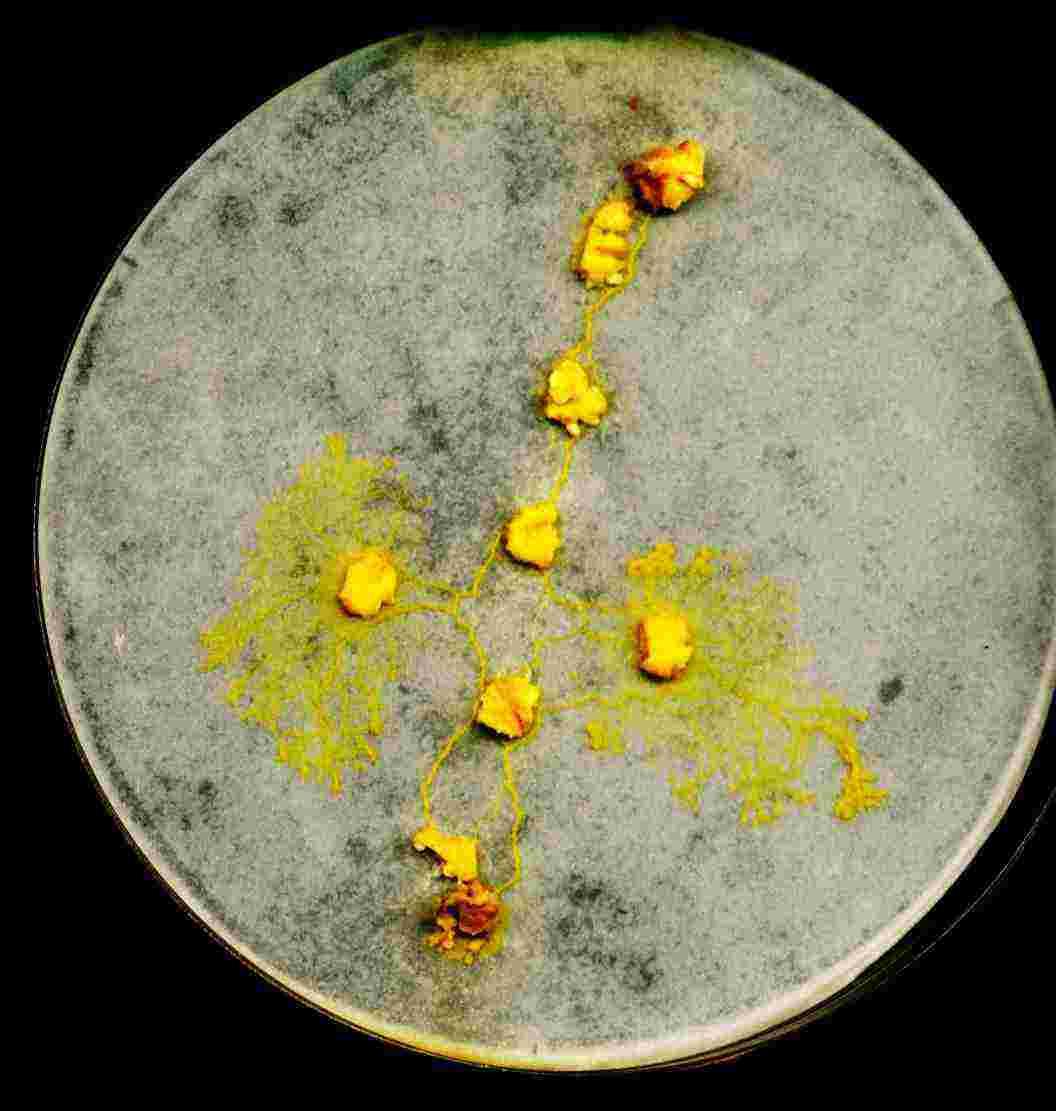}}
\caption{Multiplication of active zone.}
\label{multiplication}
\end{figure}

Multiplication of an active zone, the operation {\sc Mult}$(A)=\{A_1, A_2\}$, can be implemented by placing sources of nutrients
nearby the protoplasmic tubes, or inactive zones (Fig.~\ref{multiplication}). An example is shown 
in Fig.~\ref{multiplication}. A chain of oat flakes is connected by protoplasmic veins, 
we add two new oat flakes to evoke new active zones (Fig.~\ref{multiplication}a). Ten hours later two active 
zones $A_1$ and $A_2$  are formed, each pseudopodium travels to its unique oat 
flake (Fig.~\ref{multiplication}b).

\begin{figure}
\centering
\subfigure[$t=63$~steps]{\includegraphics[width=0.4\textwidth]{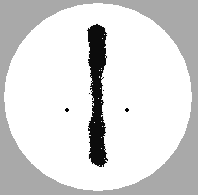}}
\subfigure[$t=126$~steps]{\includegraphics[width=0.4\textwidth]{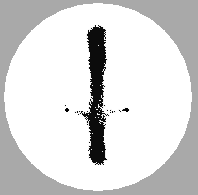}}
\subfigure[$t=147$~steps]{\includegraphics[width=0.4\textwidth]{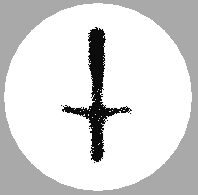}}
\subfigure[$t=273$~steps]{\includegraphics[width=0.4\textwidth]{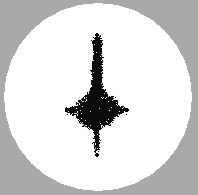}}
\caption{Evolution of model, simulating multiplication of active zones. }
\label{multiplicationModel}
\end{figure}

The model output initially recreates the spanning tree from the outer nodes, using the method in (Fig.~\ref{fusingModel}). Two more food sources are added to either side of the array and the plasmodium sends two streams outwards to engulf the sources (Fig.~\ref{multiplicationModel}). When the food sources have been engulfed, the plasmodium shifts in position by redistributing its component parts to cover the (diamond shaped) area created by the addition of the two new nodes.

\begin{figure}
\centering
\subfigure[$t=$0~h]{\includegraphics[width=0.4\textwidth]{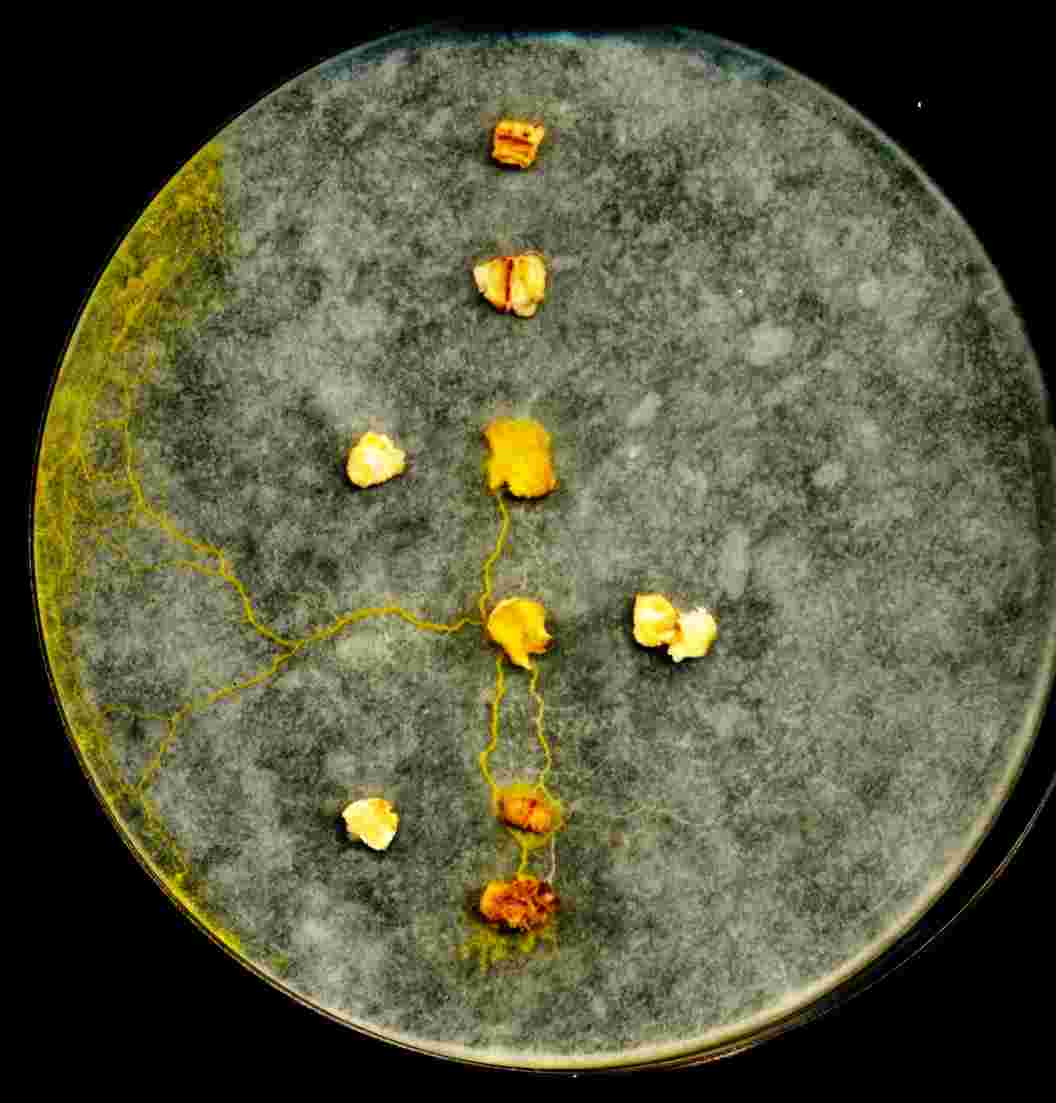}}
\subfigure[$t=$12~h]{\includegraphics[width=0.4\textwidth]{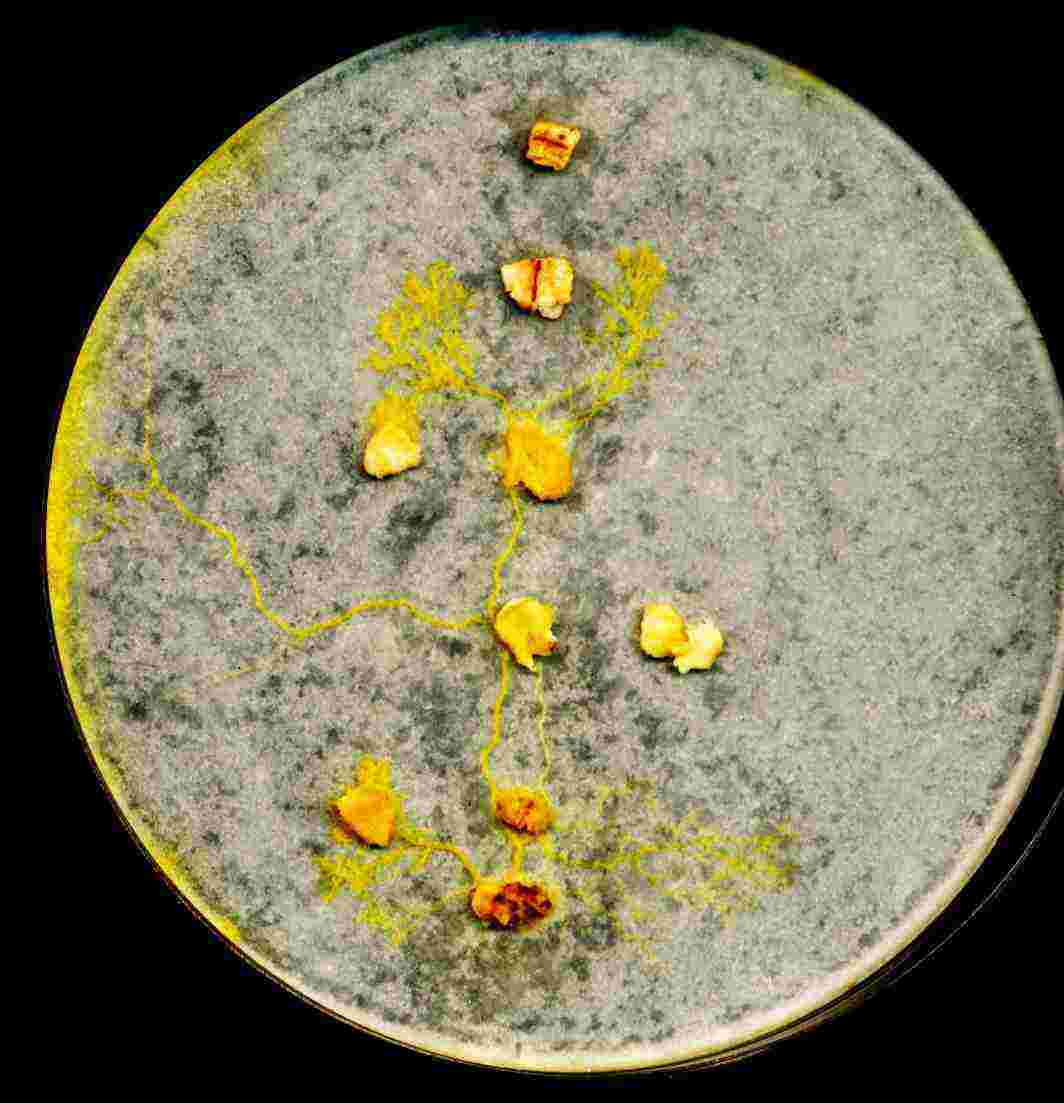}}
\caption{Illustration of how active zone can be relocated to a domain without food sources, 
operation {\sc Direct}$(A, {\bf v} )$.}
\label{directionbytwooatflakes}
\end{figure}

In some case we may need to translate active zone not to another graph node (source of nutrients) but
to a domain of an active space. Moreover, it may be necessary to provide the active zone  $A$ with certain 
initial velocity vector $\bf v$ to the zone continues its propagation in the predetermined direction.
Such operation {\sc Direct}$(A, {\bf v} )$ is executed using additional sources of 
nutrients (Fig.~\ref{directionbytwooatflakes}). Given a chain of oat flakes connected by protoplasmic 
tubes we added two new flakes on the west of the chain (Fig.~\ref{directionbytwooatflakes}a). An active 
pseudopodium, representing the active zone $A$  sprouted from the `old' flake lying between projections of the new flakes. 
It continues propagating along the bisector separating two new flakes. The active zone continues its propagation 
towards West till collide with the wall of the Petri dish. Meantime, two more active zones are formed to 
connect new oat flakes to the existing protoplasmic network (Fig.~\ref{directionbytwooatflakes}b). 

The model was not able replicate the experimental results of Fig.~\ref{directionbytwooatflakes}. As soon as a food stimulus is removed from the environment the model plasmodium begins to shrink. We speculate that the agent population is responding too quickly to the changing environment, and so the inertia of the agent population (in the current version of the model) is not as strong as observed in the real plasmodium. The model of Physarum machine requires some direct stimulus to move the plasmodium active zone. If it is not possible to use food source stimuli to move the active zone then it may be possible to use a synthetic hazard (such as the response of real plasmodia to certain frequencies of visible light) to drive the active front to a new position by repulsion.

\begin{figure}
\centering

\subfigure[$t=$0~h]{\includegraphics[width=0.4\textwidth]{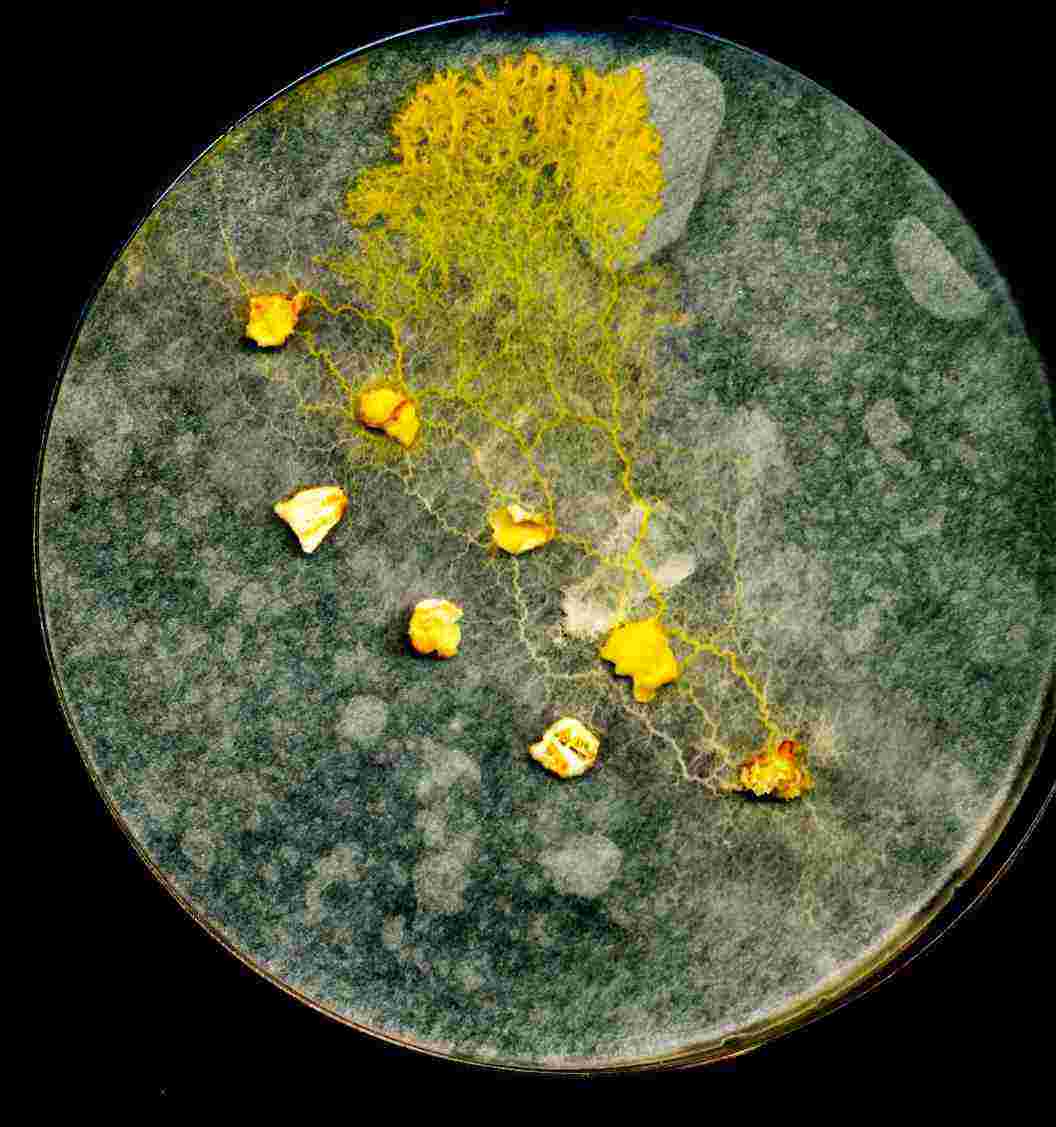}}
\subfigure[$t=$11~h]{\includegraphics[width=0.4\textwidth]{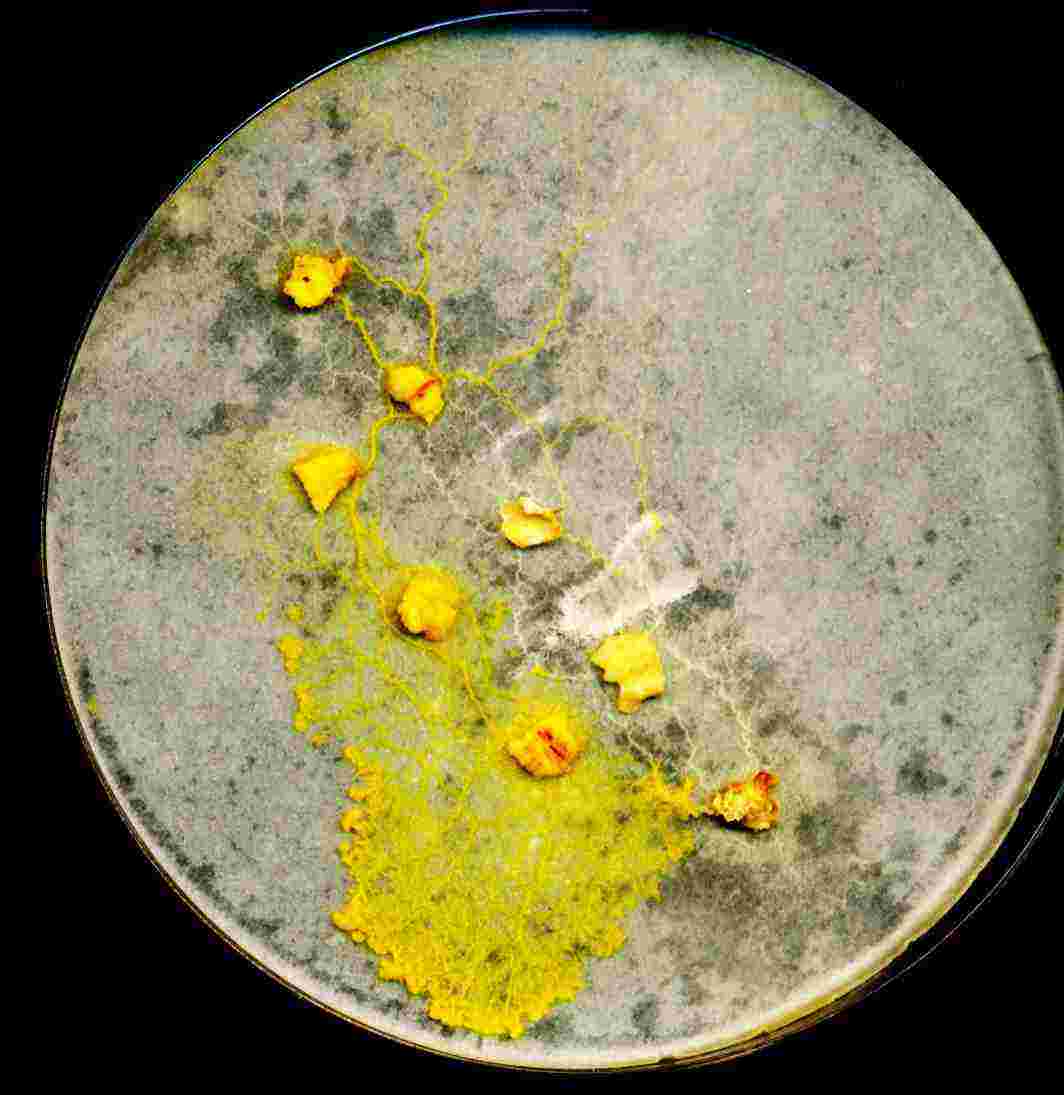}}
\caption{Major restructuring of the graph. }
\label{threenodeshift}
\end{figure}
 
Translation of active zone may lead to, or used for, undertaking of the major restructuring of the data storage graph. 
Thus, a parallel partial shift of the graph chain is demonstrated in Fig.~\ref{threenodeshift}.
Three new oat flakes are places alongside the chain of oat flakes already connected
by protoplasmic tubes (Fig.~\ref{threenodeshift}a). In few hours the plasmodium occupies new oat flakes. 
Moreover, part of the protoplasmic graph coaligned with new oat flakes is shifted to new oat 
flakes (Fig.~\ref{threenodeshift}b). Abandoned protoplasmic veins, former edges of the shift part, 
are visible as white tubes.

Such a major restructuring of the graph was caused by relocation of the active zone. At first the active zone travelled 
North-East (Fig.~\ref{threenodeshift}a). Addition of new oat flakes caused the zone to switch to the new location and 
move South-West (Fig.~\ref{threenodeshift}b).

\begin{figure}
\centering
\subfigure[$t$=1372, 1428, 1484, 1568, 1820 steps]{\includegraphics[width=0.95\textwidth]{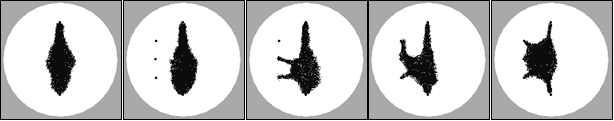}}
\subfigure[$t$=2016, 2156, 2212, 2296, 2352 steps]{\includegraphics[width=0.95\textwidth]{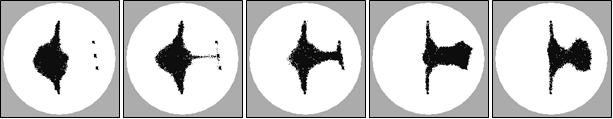}}
\caption{Major restructuring of the graph in the particle model. 
(a)~Active front shifts towards new stimuli on the left.
(b)~Active front shifts towards new stimuli on the right. }
\label{threenodeshiftModel}
\end{figure}

A major restructuring was also observed in experiments with the particle model. Note that the particle model exhibits distinct differences to the actual organism. The graph restructuring with the real plasmodium is affected by the remnants of old protoplasmic veins. Such debris is not present in the model (primarily since the `veins' in the model merely consist of protoplasmic flow itself and have no structural features). The shift in active front position is dramatically illustrated in Fig.~\ref{threenodeshiftModel}a where the front moves from its central position to engulf the three new nodes on the left. Fig.~\ref{threenodeshiftModel}b continues the same experiment, deleting the three leftmost nodes and adding three new right-side nodes, resulting in a significant surge of the active front to the right side of the `dish'.

In many examples above we observed formation of protoplasmic tubes, connecting two geographically closest
sources of nutrients. Given a chain of oat flakes, an additional oat flake is added to the experimental 
container. What is exact mechanism of inclusion the new flake in the graph? Do all nodes of the graph 
develop active zones, which travel to the flake while competition with each other for the flake? 

Our experimental observations show that 
if the graph remains connected then only one active zone, heading for 
the new oat flake, emerges. This happens due to sychronization 
of activity in the whole protoplasmic network.

This may indicate that plasmodium first `decides' which part of the protoplasmic graph closest to the recently 
added source of nutrients and only then generates an active zone. All parts of the Physarum machine sense the
chemo-attractants coming from the new source of nutrients, however the node closest to the new source somehow 
suppresses, or inhibits, activity of other nodes. Existence of connections between nodes of the Physarum machine 
is a pre-requisite for the inhibition. 

\begin{figure}
\centering
\subfigure[$t$=0~h]{\includegraphics[width=0.4\textwidth]{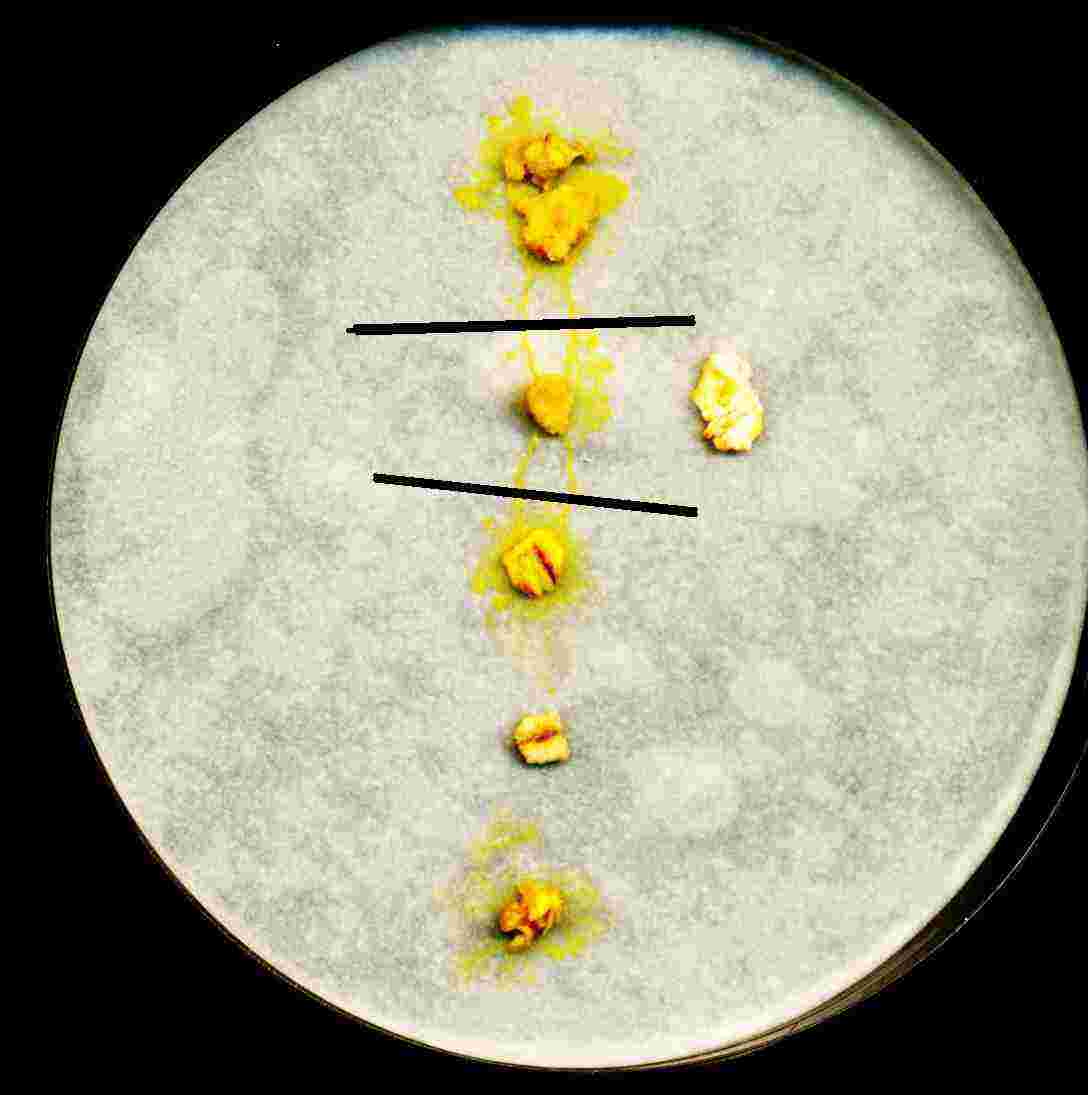}}
\subfigure[$t$=11~h]{\includegraphics[width=0.4\textwidth]{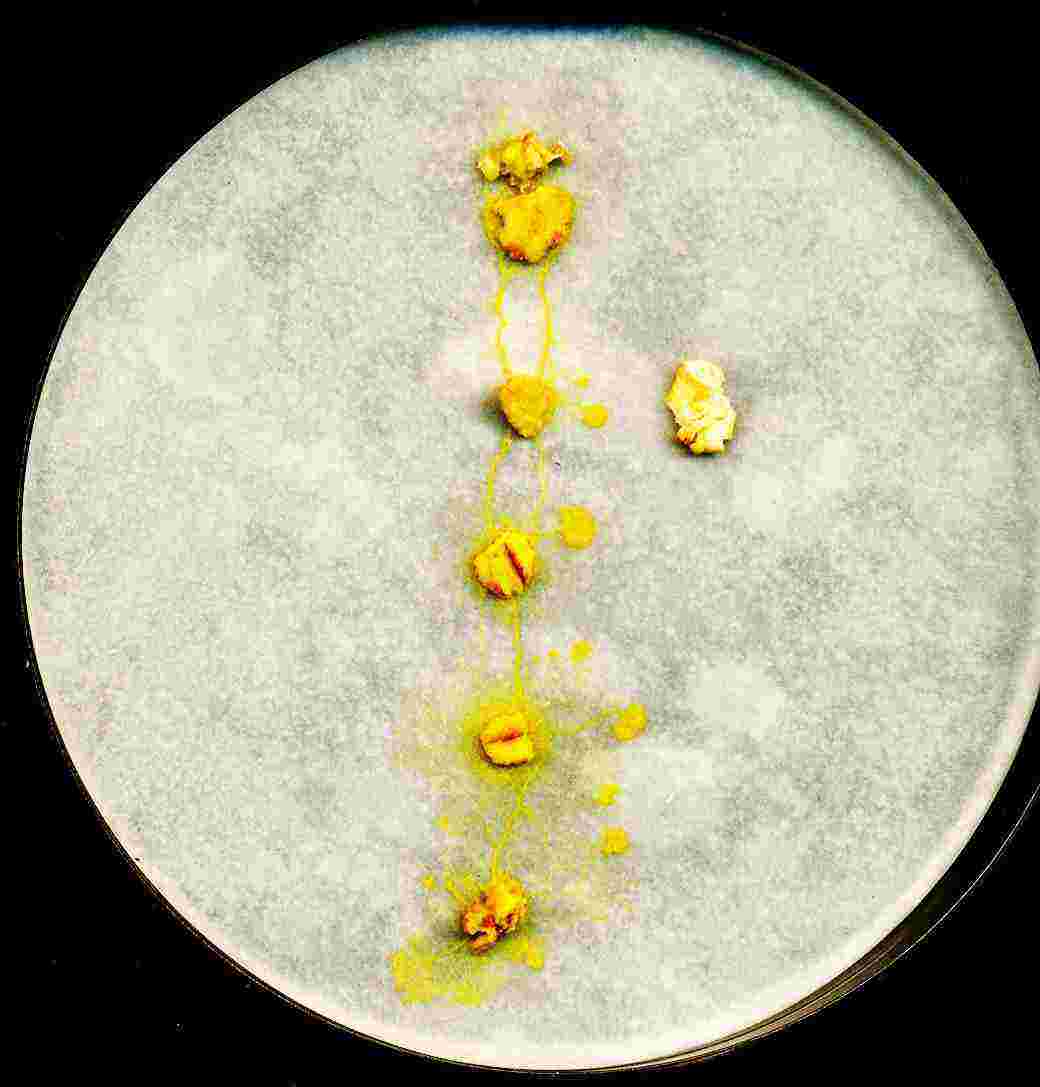}}
\subfigure[$t$=17~h]{\includegraphics[width=0.4\textwidth]{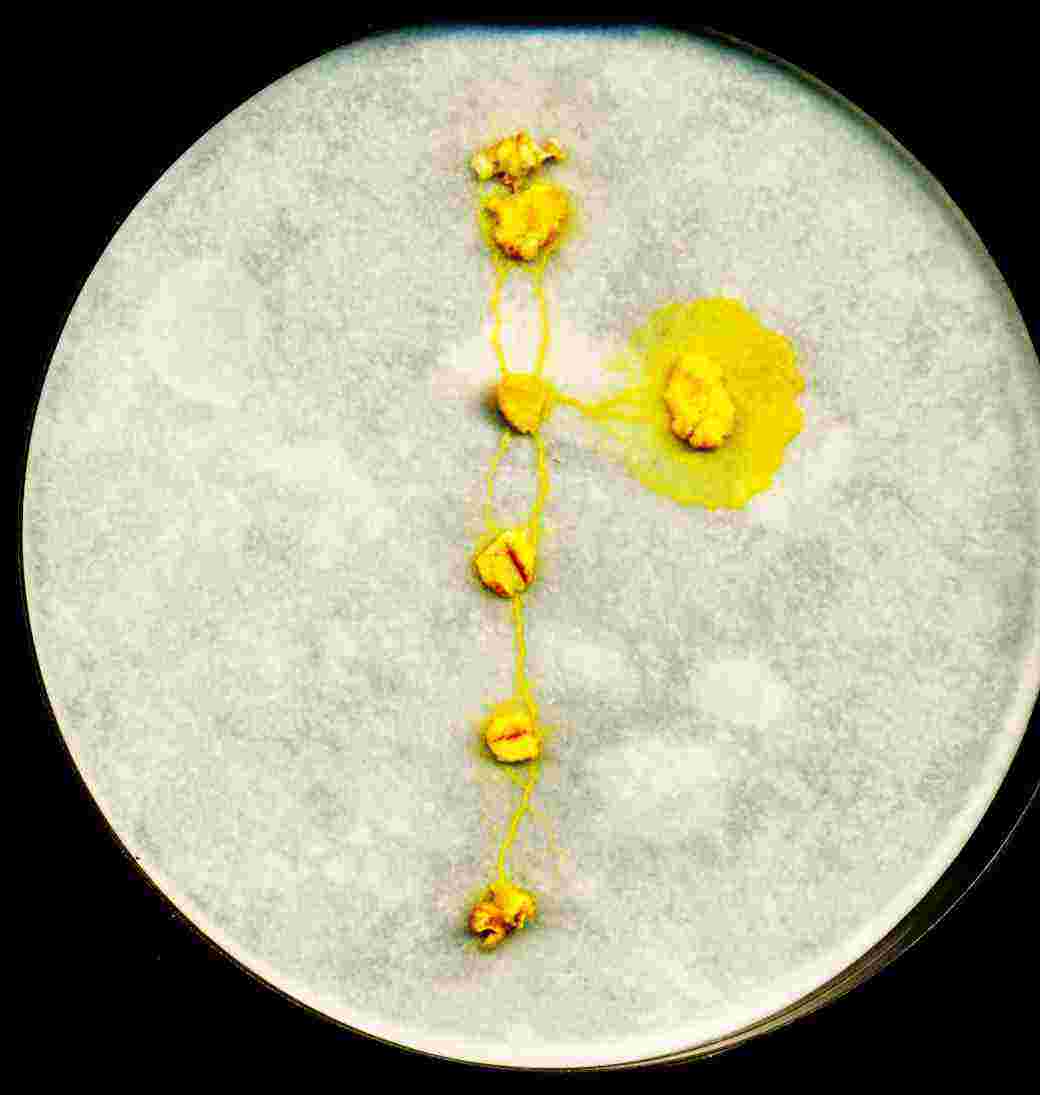}}
\caption{Cutting edge leads up to de-synchronization of the Physarum machine.}
\label{desynchro}
\end{figure}

To prove the point we arranged oat flakes in a chain and allowed the chain to be spanned by the plasmodium~(Fig.~\ref{desynchro}a). 
When the flakes are connected by protoplasmic tubes we add new oat flake, the eastmost flake on the picture~(Fig.~\ref{desynchro}a). 
To break communication between nodes of the Physarum machine we cut through protoplasmic tubes connecting three Northern nodes, 
cut places are shown by lines in~Fig.~\ref{desynchro}a). Due to breakup in communication all nodes of the plasmodium 
reacts, almost simultaneously, to the addition of new source of nutrients by sprouting pseudopodia~(Fig.~\ref{desynchro}b).
In few hours cut tubes are restored~(Fig.~\ref{desynchro}c). This reinstates communication between all parts of the Physarum machine.
The distant, from the new flake, node of the Physarum machine des-activate their active zones (cease propagation of pesudopodia).
In the result of restored communication  only one active zone remains, and the new node is connected to the storage graph by a single edge.

\begin{figure}
\centering
\subfigure[$t$=0~h]{\includegraphics[width=0.4\textwidth]{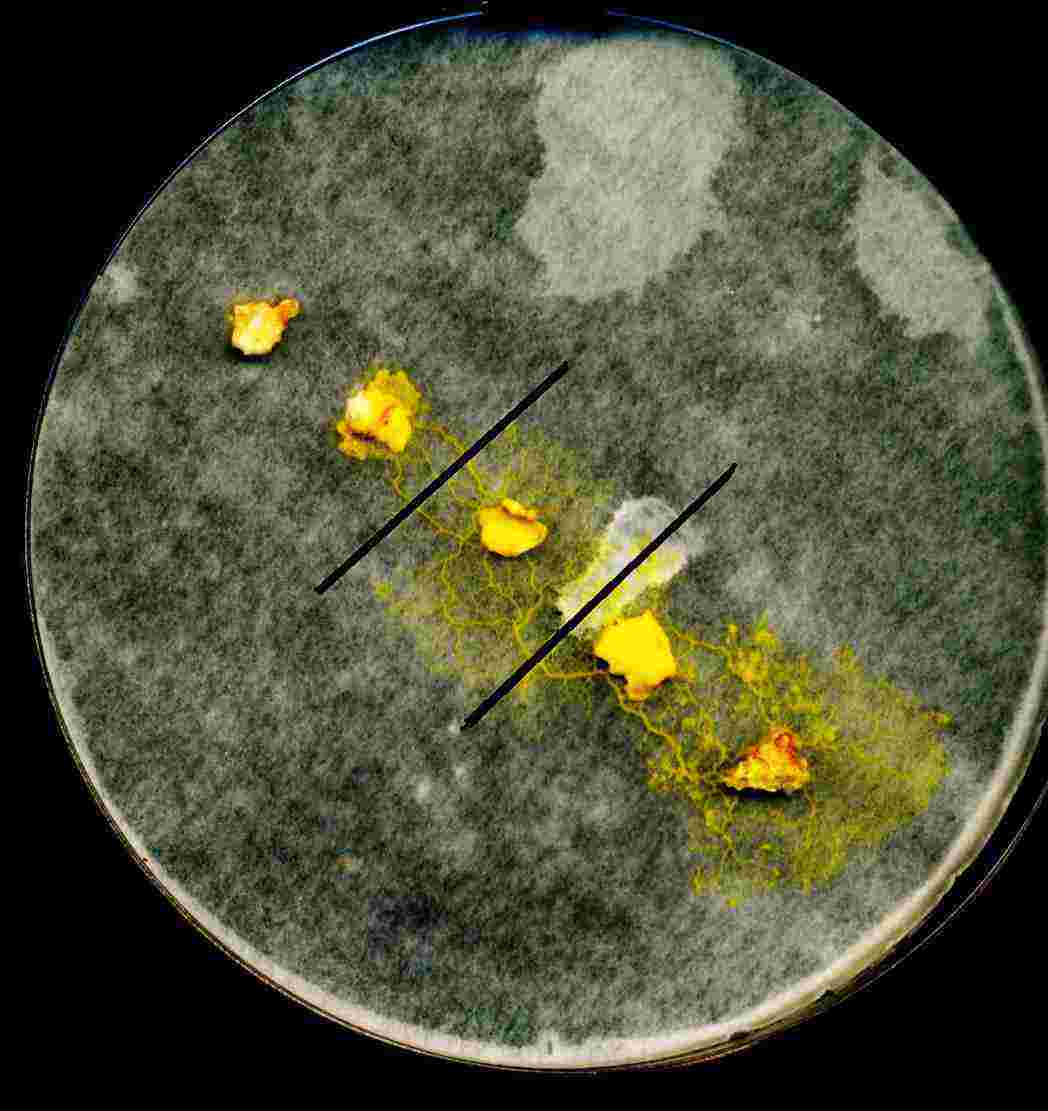}}
\subfigure[$t$=14~h]{\includegraphics[width=0.4\textwidth]{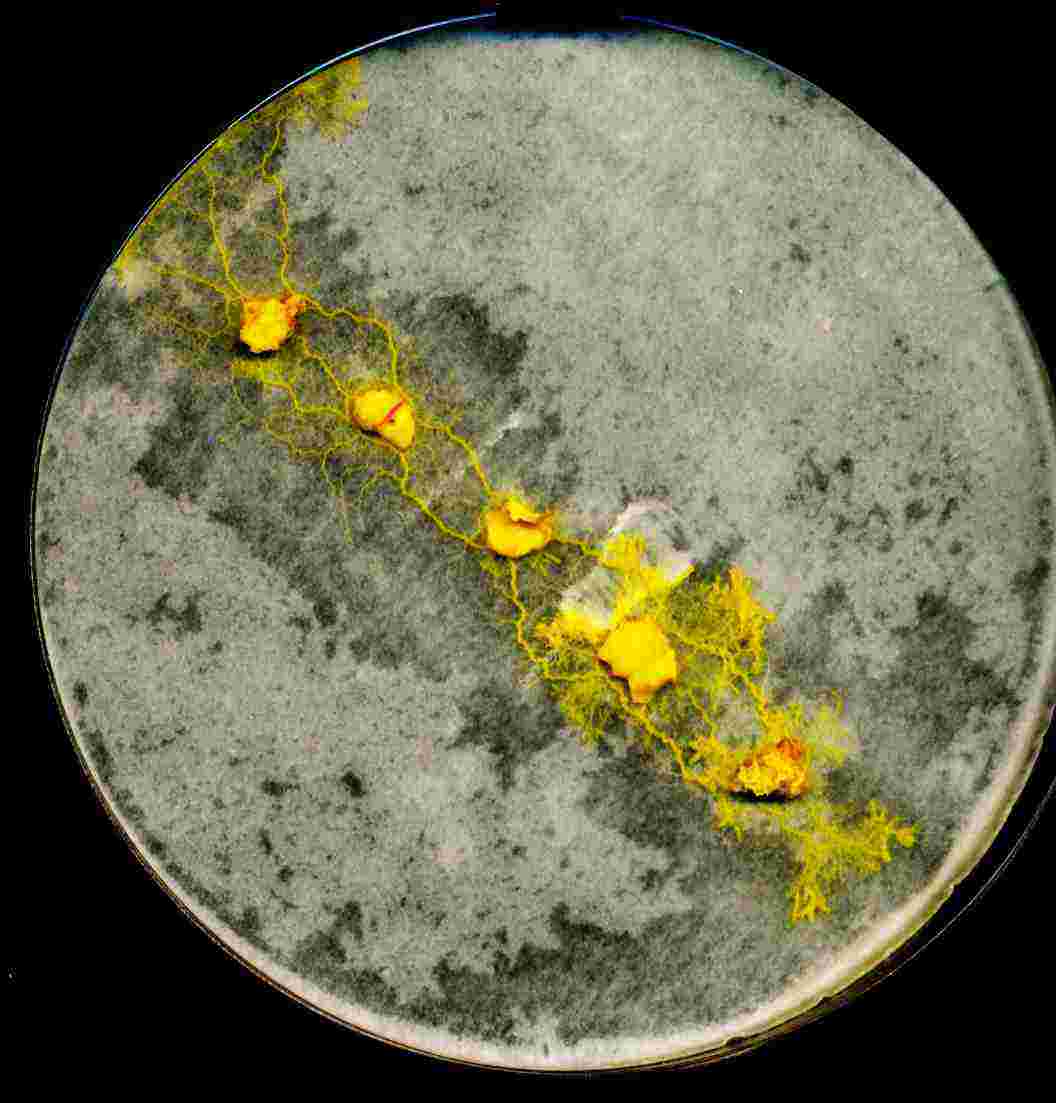}}
\caption{Absence of adverse reaction to tube cutting.}
\label{cuttingOK}
\end{figure}

Just cutting protoplasmic tubes does not lead to formation of new pseudopodia (Fig.~\ref{cuttingOK}).
When tubes are cut they do usually fuse back again in few hours without any adverse effect to the 
plasmodium behaviour.

The model plasmodium was not able to replicate the experiment of Fig.~\ref{desynchro}, once again suggesting that the model is currently lacking in the mechanism which couples the plasmodium behaviour across food sources.

\section{Summary}
\label{discussion}

We studied the Physarum machine, an implementation of general-purpose Kolmogorov-Uspensky machine (KUM) in 
vegetative stage, plasmodium, of \emph{ Physarum polycephalum}. Our present paper is mainly concerned with 
manipulating active zones (actively growing pseudopodia of the plasmodium) because they are main computational
units of the storage modification machine, implemented in the plasmodium. In laboratory experiments and computer 
simulation of the plasmodium we executed basic operations with active zones. We shown 
how merge two active zones, to multiply an active zone, to translate active zone to a new node of the storage structure, 
to direct active zone in the space not occupied by data nodes. We envisage our experimental and theoretical findings 
will be employed in future programming of spatially distributed biological computers.

\end{document}